# Vectorized Context-Aware Embeddings for GAT-Based Collaborative Filtering

Danial Ebrat [†, *], Sepideh Ahmadian [†, *], Luis Rueda [†, *]

† School of Computer Science, University of Windsor

**Abstract**

Recommender systems often struggle with data sparsity and cold-start scenarios, limiting their ability to provide accurate suggestions for new or infrequent users. This paper presents a Graph Attention Network (GAT)-based Collaborative Filtering (CF) framework enhanced with Large Language Model (LLM)-driven context-aware embeddings. Specifically, we generate concise textual user profiles and unify item metadata (titles, genres, overviews) into rich textual embeddings, injecting these as initial node features in a bipartite user–item graph. To further optimize ranking performance, we introduce a hybrid loss function that combines Bayesian Personalized Ranking (BPR) with a cosine similarity term and robust negative sampling, ensuring explicit negative feedback is distinguished from unobserved data. Experiments on the MovieLens 100k and 1M datasets show consistent improvements over state-of-the-art baselines in Precision, NDCG, and MAP while demonstrating robustness for users with limited interaction history. Ablation studies confirm the critical role of LLM-augmented embeddings and the cosine similarity term in capturing nuanced semantic relationships. Our approach effectively mitigates sparsity and cold-start limitations by integrating LLM-derived contextual understanding into graph-based architectures. Future directions include balancing recommendation accuracy with coverage and diversity, and introducing fairness-aware constraints and interpretability features to enhance system performance further.

## 1. Introduction

Recommender systems are critical for personalized content delivery in domains like e-commerce, streaming, and social media, analyzing user preferences to suggest relevant items and improve engagement. Traditional collaborative filtering (CF) methods, relying on user-item interaction matrices, effectively capture latent patterns but face challenges such as data sparsity, cold-start problems, and limited contextual integration.

To address these issues, Matrix Factorization (MF) techniques such as Singular Value Decomposition (SVD) and Alternating Least Squares (ALS) [1, 2] have been employed, improving accuracy but still struggling with sparsity and contextual richness. This has spurred the integration of side information, such as item content, social networks, and knowledge graphs, to enhance CF performance [3, 4].

Graph-based CF methods have emerged as a promising alternative, leveraging graph structures to model user-item interactions more effectively. Early approaches, such as ItemRank [5] and BiRank [6], used label propagation but lacked optimization capabilities. More advanced techniques, like HOP-Rec [7], integrated random walks with BPR. However, these models remain highly sensitive to hyperparameter tuning and often fail to capture high-order collaborative signals effectively.

Graph Neural Networks (GNNs) have revolutionized recommendation systems by capturing complex user-item interactions, particularly in sparse data scenarios. Models like GC-MC[8] and PinSage[9] enhance user-item and item-item relationships, while SpectralCF[10] leverages spectral convolutions but faces scalability challenges. Among GNN-based CF approaches, Neural Graph Collaborative Filtering (NGCF) [11] has gained attention for explicitly encoding high-order connectivity through iterative message-passing. While NGCF improves recommendation accuracy, its non-linear transformations increase computational complexity and risk of overfitting. LightGCN [12] addresses these issues by simplifying the Graph Convolutional Network (GCN), focusing on the weighted aggregation of neighborhood information, and preserving essential collaborative signals efficiently, making it a strong baseline for graph-based recommender systems.

* ebrat@uwindsor.ca

Self-attention mechanisms, inspired by the Transformer architecture, have been widely adopted in recommender systems. BERT4Rec [13] leverages BERT [14] for sequence-based recommendations, capturing bidirectional user-item interactions, while DSIN [15] employs self-attention to model long-term user interests, improving accuracy in sequential tasks. These methods demonstrate self-attention's effectiveness in modeling complex user behavior. Beyond sequential models, knowledge graph-enhanced approaches like IGAT[16] and TKGAT [17] integrate Graph Attention Networks (GAT) to enrich user and item representations, with TKGAT incorporating tag-aware recommendations. However, scalability, noise accumulation, and interpretability remain key challenges, underscoring the need for more efficient and robust attention-based systems.

In the LLM era, recent works have integrated textual embeddings from item descriptions and user reviews using pre-trained language models (PLMs) like BERT [14] and LLMs combined with CF and graph-based approaches [18]. For instance, Lusifer [19] generates context-aware user profiles from interaction histories to enhance recommender performance, while A-LLMRec [20] combines a pre-trained CF model with an LLM to leverage collaborative knowledge directly. Although effective in cold-start scenarios, A-LLMRec underperforms in warm-start cases due to its heavy reliance on textual embeddings. These advancements demonstrate the growing synergy between CF and language understanding, but challenges remain in achieving robust performance across diverse scenarios.

Despite notable progress in collaborative filtering methods, existing solutions often struggle with data sparsity and cold-start scenarios. This paper addresses these challenges by LLM-driven context-aware embeddings within a GAT framework. Specifically, we generate concise user preference profiles and unify item metadata into rich textual embeddings, yielding more expressive node representations. We introduce a custom loss function combining BPR with a cosine similarity term to enhance ranking quality and embedding alignment. A robust negative sampling strategy additionally ensures that explicit negative feedback is effectively distinguished from unobserved data. The key contributions of this work are as follows:

1. LLM-Augmented User and Item Representations: We enhance user and item nodes with context-aware embeddings generated from concise textual profiles and item metadata, which are injected as initial node embeddings during training.
2. Hybrid Loss Function: We introduce a combined loss function that integrates BPR with a cosine similarity term and robust negative sampling to optimize ranking performance and the alignment of semantically similar embeddings.

## 2. Methodology

This section illustrates the operational procedure of our methodology, as illustrated in Figure 1, and for transparency and reproducibility, the implementation details are available in our GitHub repository[1] . We use the MovieLens 100k and 1m datasets [21] and enrich both datasets with metadata from The Movie Database (TMDB) API by retrieving genres and movie overviews and concatenating titles, genres, and overviews into a single textual feature. This unified text representation preserves semantic relationships within the text more effectively than averaging separate embeddings (e.g., title, genre), which can dilute meaningful information. It also enables the model to weigh different textual elements dynamically based on context.

---

[1] https://github.com/danialebrat/ClusteredKNN_NMF

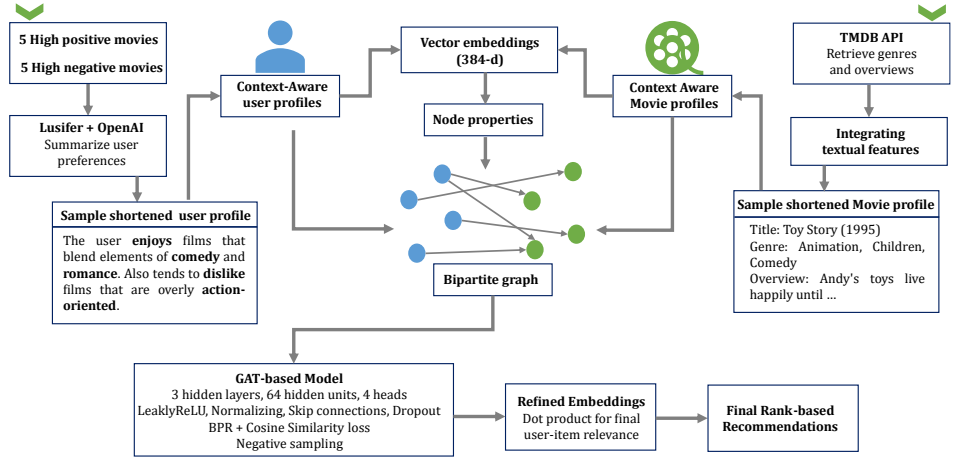

*Figure 1. Methodology pipeline: pre-processing and training procedure.*

In practice, extreme scores carry substantially stronger and less ambiguous sentiment than mid-range opinions. Therefore, for every user, we retain only the top 5 items that were loved (r = 5), and strongly disliked (r ≤ 2), excluding mid-range ratings for more apparent distinction. Based on our initial experiments, fewer ratings under-represent the breadth of a user's taste, and higher numbers often repeat information or confuse the LLM. We employ Lusifer [19] and the OpenAI 4o-mini model to generate concise user profile textual features, summarizing key descriptors such as genres and notable plot details. These profiles approximate real users' opinions by aligning closely with positively rated movies and diverging from negatively rated ones.

We encode each movie's concatenated text and each user's profile text into 384-dimensional vectors using the all-MiniLM-L6-v2 model [22]. These embeddings serve as initial node features in a bipartite user–item graph, where edges represent explicit ratings. Treating edges as bidirectional enables message passing in both directions, ensuring mutual influence between user and item embeddings.

We adopt a GAT to refine embeddings by aggregating information from neighboring nodes. Specifically, we stack three GAT layers with 64 hidden units and four attention heads. Each layer employs LeakyReLU activation, layer normalization, skip connections, and dropout to improve training stability and mitigate overfitting. For optimization, we combine the BPR loss [23] with a cosine similarity term to align user and item embeddings in positive interactions. The overall loss is: $Loss = BPR\ Loss + \alpha \times (1 - Cosine\ Similarity)$, where α balances the ranking and embedding-alignment terms.

Ratings of 4 or 5 are treated as positive, and 1 or 2 as negative (excluding rating 3 and unrated items). We use the AdamW optimizer with weight decay, an adaptive learning rate (initially with 0.001 and reduced by a factor of 0.4 after five steps of patience), and an early stopping strategy. After training converges, we obtain refined user and item embeddings; their dot product provides the final relevance score for each user-item pair.

### 3. Experimental Results

We partitioned datasets into five folds for 5-fold cross-validation with an 80:20 train–test split. Given the high sparsity in MovieLens (approximately 97% of possible user-item interactions are missing), we only evaluated methods on the items with explicit user ratings, thereby focusing on measured relevance rather than penalizing the model for unobserved items that may still be relevant. We used three hidden layers with 64 hidden units, with the same final node embedding size for all graph-based models to ensure the comparison's fairness. We evaluated our approach using various metrics and compared it with Neural Graph Collaborative Filtering (NGCF), LightGCN, Alternating Least Squares (ALS), and an ablated GAT lacking LLM-based context-aware embeddings and cosine similarity term in the loss function. The results, shown in Figure 1, are averaged over five cross-validation folds.

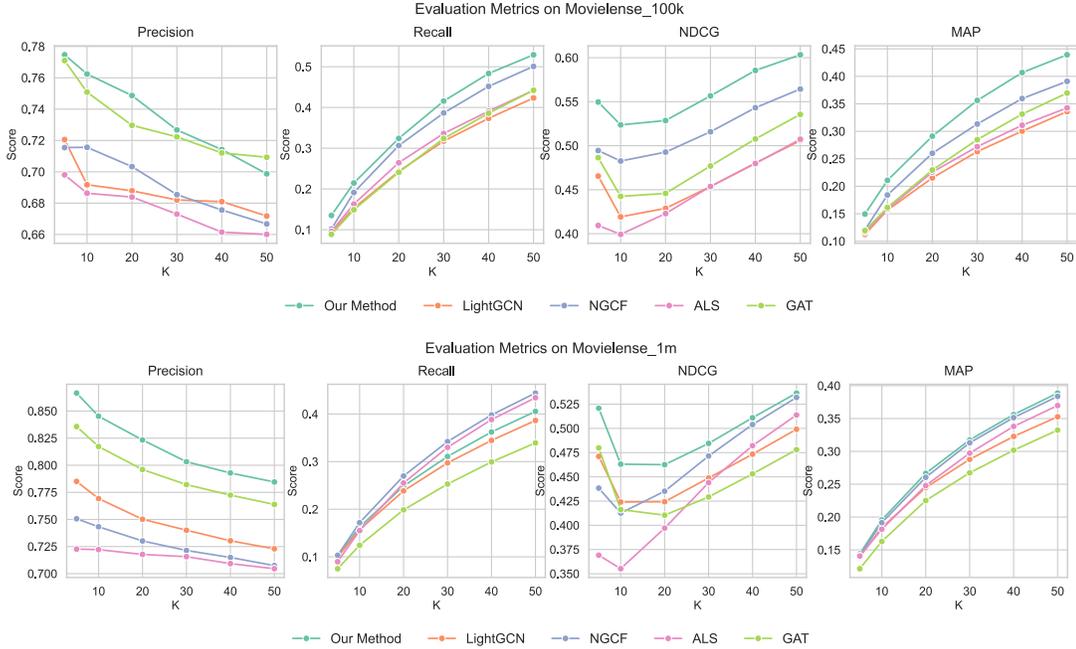

*Figure 2. Comparison of metrics using various K values (top K recommendations)*

Across both the 100k and 1M datasets, our method consistently outperforms all baselines on key ranking metrics such as Precision, NDCG, and MAP. Although there is a marginal decrease in Recall on the 1M dataset compared to NGCF, the gain in other metrics suggests the proposed method emphasizes correctly ranking truly relevant items higher in the recommendation list. Precision, NDCG, and MAP improvements can be attributed to two key components of our approach. First, using LLM-based user profiles enables the generation of concise yet salient summaries of user preferences, capturing nuanced aspects such as favored genres, typical plot elements, or stylistic features. These summaries enhance the alignment between user and item embeddings, leading to more accurate recommendations. Second, incorporating a cosine similarity term in the objective function explicitly encourages user and item embeddings to be closer to positive interactions. This alignment results in more discriminative embeddings, as evidenced by the higher ranking metrics.

On the other hand, the ablated GAT (using Xavier initialization for random embeddings as default in place of LLM-based text features and without the cosine similarity term) shows notably lower performance. This decline highlights that neither default random embeddings nor the BPR loss alone can sufficiently capture the content-driven relationships between users and items in a highly sparse environment.

To assess robustness in limited-interaction scenarios, we evaluated the methods on users with fewer than five rated items in the MovieLens 100k dataset (Table 1). The 1m dataset doesn't include a cold start scenario, as the minimum interaction is 13 with only a few users. Despite having less historical data, our method consistently outperforms the baselines across most ranking metrics and improves item coverage. LLM-based user profiles likely contribute to this resilience; even with minimal training examples, summarizing user interests via textual cues offers an informative prior that aligns new or infrequent users with content that reflects their expressed preferences.

Overall, these experiments underscore the effectiveness of integrating LLM-based textual representations within a graph-based recommender architecture. We obtain more expressive node features by summarizing user preferences and item metadata as textual embeddings. Adding a cosine similarity term further refines the latent space, leading to enhanced alignment of user and item nodes for highly relevant matches. At the same time, the emphasis on semantic context can skew recommendations toward items with richer descriptions, impacting diversity metrics.

Looking ahead, balancing accuracy with coverage and diversity remains a key challenge. Introducing techniques such as regularization for novelty or fairness-aware constraints and bringing interpretability and explainability could help mitigate the popularity bias.

| Method | Precision | Recall | NDCG | MAP | Item Coverage |
|---|---|---|---|---|---|
| ALS | 0.285714 | 0.060606 | 0.231657 | 0.060606 | 0.125 |
| NGCF | 0.538462 | 0.212121 | 0.419805 | 0.13035 | 0.232143 |
| LightGCN | 0.5 | 0.090909 | 0.247584 | 0.057576 | 0.107143 |
| Ablated GAT | **0.666667** | 0.060606 | 0.21306 | 0.050505 | 0.053571 |
| Our method | 0.571429 | **0.242424** | **0.456341** | **0.143778** | **0.25** |

*Table 1. Results of the top 20 recommendations considering users with less than five interactions (100K dataset).*

## 4. Conclusion and Future Work

We developed a GAT recommender system enhanced with LLM-generated textual features as initial node embeddings and a cosine similarity term in its loss function. This design captures detailed semantic relationships and yields more precise user–item embeddings than approaches relying solely on structured rating data.

Although our GAT architecture consistently outperforms traditional methods in accuracy-focused metrics, balancing high performance with fair and diverse item recommendations remains a key challenge. Future work should explore diversity- and fairness-aware strategies to ensure broader, more equitable item exposure. Additionally, enhancing explainability is crucial, particularly given the complexity of attention mechanisms and LLM-based embeddings. Potential solutions involve analyzing attention weights or applying established interpretability techniques (e.g., LIME, SHAP, or Integrated Gradients) to illuminate the impact of specific textual features and user-item interactions on recommendation outcomes.